\documentclass[aps,twocolumn,preprintnumbers]{revtex4}

\usepackage{graphicx}  % Include figure files
\usepackage{subfigure}
\usepackage{multirow}
\usepackage{natbib}
\usepackage{balance}
\usepackage{flushend}

\usepackage{fancyhdr}
\usepackage{longtable}
\usepackage{parskip}
\usepackage[T1]{fontenc}
\usepackage{dcolumn}   % Align table columns on decimal point

\usepackage{bm}        % bold math
\usepackage{amsfonts}  % Common math fonts
\usepackage{amsmath}   % Common math functions
\usepackage{amssymb}   % Common math symbols

\newcommand{\pwisein}{\left\{ \begin{array}{ll}}
\newcommand{\pwiseout}{\end{array}\right.}
\newcommand{\ket}[1]{\left| #1 \right\rangle}

\begin{document}

\title{Spontaneous creation of skyrmions in a two-component Bose-Einstein condensate}

\author{Rabeea Kiran and Hiroki Saito}

\affiliation {\it Department of Engineering Science, University of Electro-Communications, Tokyo 182-8585, Japan}

\begin{abstract}  

We investigate the stability of a vortex ring in a miscible
two-component Bose-Einstein condensate confined in a harmonic
potential, where the vortex cores in the two components are initially
overlapped.
Solving the Gross-Pitaevskii equation numerically, we find
that the overlapped vortex rings in the two components are dynamically
unstable against separation and that they can form linked vortex
rings, resulting in a three-dimensional skyrmion.
The parameter range for spontaneous skyrmion generation is determined by the Bogoliubov analysis.

\end{abstract}

\maketitle 

\section{Introduction}

Skyrmions are nontrivial topological structures proposed in
nuclear physics to explain hadrons \cite{skyrme}.
Skyrmions have since transcended their theoretical origins and have
been extended to various physical systems.
Numerous theoretical proposals have
facilitated the observation and manipulation of skyrmions in chiral
magnets \cite{m1}, liquid crystals \cite{crys}, quantum Hall systems
\cite{hall1, hall2, hall3}, superconductors \cite{super1, super2},
semiconductors \cite{solid}, acoustics \cite{acous}, superfluid
$^3{\rm He}$ \cite{flu1, flu2}, and Bose-Einstein condensates (BECs)
of ultracold gases \cite{bec1,bec2}. The order parameters with
internal degrees of freedom enable the atomic BECs to support a wide
range of topological entities such as dark solitons \cite{soliton1},
quantized vortices \cite{vortices1,vortices2}, monopoles
\cite{mono1,mono2,mono3}, and knots \cite{knots1,knots2,knots3}. The
system of a multicomponent BEC is thus a suitable platform for
examining the behavior of topological spin structures, including
skyrmions.

Various studies on skyrmions in multicomponent BECs have been reported.
Two-dimensional (2D) skyrmions have been realized
experimentally using the Raman process \cite{raman} and magnetically
induced spin rotations \cite{rot}.
Three-dimensional (3D) skyrmions have been successfully created in a
spin-1 BEC using a spin rotation technique \cite{spin-1}.
A wide variety of methods for creating
skyrmions in multicomponent BECs have been proposed, including
electromagnetically induced transitions \cite{bec1}, spin manipulation
using a structured magnetic field and laser beams
\cite{spin-2,zamora,q,imprint}, spin-orbit interaction
\cite{coupling1,coupling2}, annihilation of domain walls
\cite{domain}, Plateau-Rayleigh instability \cite{capillary}, optical
excitation of atoms \cite{optical}, and a moving obstacle
\cite{hydro}. The stability and dynamics of skyrmions have also been
investigated extensively \cite{stoof,stable1,lifetime, Simula,
  stable2,stable3,stable4,stable5,stable6,stable7,stable8,stable9,stable10}. 

In this paper, we show that a 3D skyrmion is generated spontaneously
in a two-component BEC. As an initial state, we consider a U(1) vortex
ring in a miscible two-component BEC, where quantized vortex rings in
the two components are totally overlapped with each other. These
overlapped vortex rings are shown to be dynamically unstable against
separation. For some parameters, we find that the separated vortices
are linked with each other, resulting in a skyrmion with an integer
winding number.

The remainder of the paper is organized as follows.
A skyrmion in a two-component BEC is introduced in
Sec.~\ref{s:skyrmion}.
The dynamics of the system and the spontaneous generation of skyrmions
are demonstrated in Sec.~\ref{s:numerical}.
The Bogoliubov analysis to find the parameter range for the skyrmion
generation is described in Sec.~\ref{s:Bogo}.
An experimental scenario to observe the formation of a 3D skyrmion is
proposed in Sec.~\ref{s:exp}, and conclusions drawn from this study
are offered in Sec.~\ref{s:conc}.

\section{Three-Dimensional Skyrmion in a Two-Component BEC}
\label{s:skyrmion}

We consider a two-component BEC at zero temperature, represented by the macroscopic wave functions $\Phi_{1}(\bm{r}, t)$ and $\Phi_{2}(\bm{r}, t)$ within the mean-field approximation. These two-component wave functions can be expressed as 
\begin{equation}
\bm{\Psi(r)} = \begin{pmatrix} \Phi_{1}(\bm{r})\\\Phi_{2}(\bm{r}) \end{pmatrix} = \sqrt{\rho(\bm{r})}\begin{pmatrix} \Lambda_{1}(\bm{r})\\\Lambda_{2}(\bm{r}) \end{pmatrix},
\end{equation}
where $\rho(\bm{r})$ = $|\Phi_{1}(\bm{r})|^2$ + $|\Phi_{2}(\bm{r})|^2$
is the total density and $(\Lambda_{1},\Lambda_{2})$ satisfies the
normalization condition
$|\Lambda_{1}(\bm{r})|^2+|\Lambda_{2}(\bm{r})|^2 = 1$.
The two complex numbers $\Lambda_1(\bm{r})$ and $\Lambda_2(\bm{r})$
represent a pseudospin 1/2 state on the SU(2) manifold. In a
two-component BEC, a skyrmion is characterized as a configuration in
which the physical space $\bm{r}$ is continuously and topologically
mapped onto the SU(2) manifold.
In this mapping, all the points at infinity ($|\bm{r}| \rightarrow
\infty$) are mapped onto the common state $(\Lambda_1, \Lambda_2)$;
therefore, the structure of this mapping is mathematically described
by the third homotopy group, $\pi_{3}[{\rm SU}(2)]=\mathbb{Z}$
\cite{skyrme, bec2, stoof, top}.
This mapping is classified by an integer topological invariant known
as the winding number or topological charge, which quantifies its
topological nature.

\begin{figure}[h]
\centering
\includegraphics[width=3.4in]{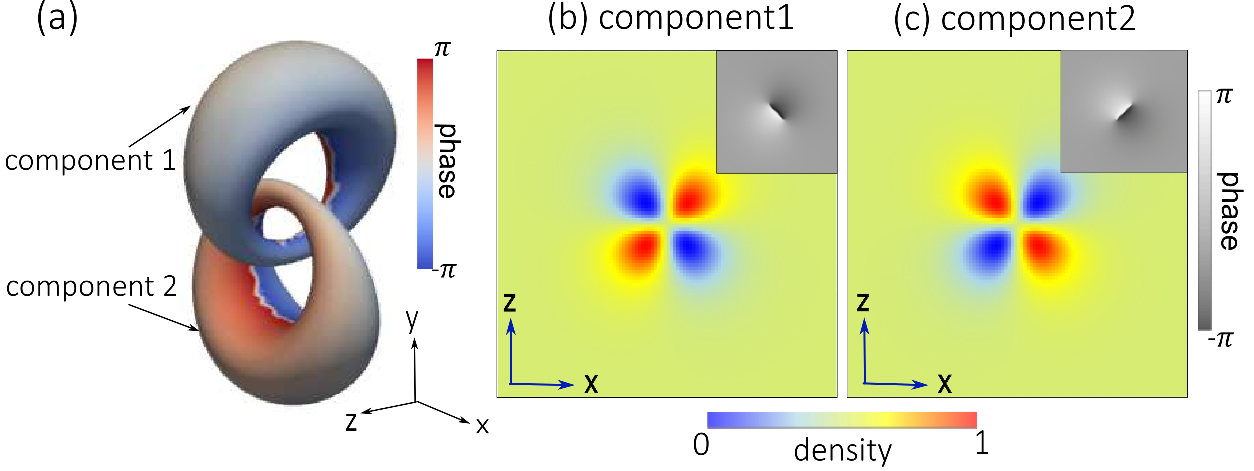}
\caption{Skyrmion state defined in Eq.~\eqref{eq:skyrmion}, which
  consists of linked vortex rings in two components.
  (a) Isodensity surface for both components ($|\psi_{1}|^{2}$ =
  $|\psi_{2}|^{2} = 0.1$); the color represents the phase at the
  surface, where the phase changes by $2\pi$ around the tubes.
  (b) and (c) Cross-sectional density (main panels) and phase (insets)
  profiles for components 1 and 2 on the $y=0$ plane,
  respectively.} \label{fig:link}
\end{figure}

A skyrmion in an equally mixed two-component BEC can be constructed by
the following expression:
\begin{widetext}
\begin{align}
\bm{\Psi(r)} = \frac{1}{\sqrt{2}}e^{-i f(r) \hat{\bm{r}} \cdot \bm{\sigma}}\begin{pmatrix} 1\\1 \end{pmatrix} =\frac{1}{\sqrt{2}}\begin{pmatrix} \cos{f}(r) - i(\cos{\theta} + \sin{\theta}e^{-i\phi})\sin{f}(r)\\[4pt]\cos{f}(r) + i(\cos{\theta} - \sin{\theta}e^{i\phi})\sin{f}(r) \end{pmatrix}, \label{eq:skyrmion}
\end{align}
\end{widetext}
where $f(r)$ is a monotonically decreasing function satisfying the boundary condition \textit{f}(0)=$\pi$ and \textit{f}($\infty$)=0, $\bm{\sigma}$ represents the vector of the Pauli matrices, and $\hat{\bm{r}}$=($\sin{\theta}\cos{\phi}$, $\sin{\theta}\sin{\phi}$, $\cos{\theta}$) is the unit vector in polar coordinates $(r, \theta, \phi)$. The rotation matrix in Eq.~(\ref{eq:skyrmion}) covers whole elements of SU(2); hence it corresponds to a mapping of a 3D unit sphere $S^3$ parametrized by $(f(r), \theta, \phi)$ onto the SU(2) manifold. Figure~\ref{fig:link} shows the skyrmion state defined in Eq.~(\ref{eq:skyrmion}). This state contains quantized vortex rings in both components, which are linked with each other \cite{bar1}, as shown in Fig.~\ref{fig:link}.

In general, a two-component state can be written as
\begin{equation}
\bm{\Psi(r)} = \sqrt{\rho(\bm{r})}\begin{pmatrix} e^{i\eta_{1}(\bm{r})}\cos{\frac{\kappa(\bm{r})}{2}}\\[4pt]e^{i\eta_{2}(\bm{r})}\sin{\frac{\kappa(\bm{r})}{2}} \end{pmatrix},\label{eq:formula}
\end{equation}
where $\eta$ and $\kappa$ are real functions.
The winding number $\textit{W}$ for a 3D skyrmion is defined as~\cite{skyrme, bec1, bec2, stable1, stoof}
\begin{equation}
W = \frac{1}{8\pi^2}\int d\bm{r} \sum_{\alpha, \beta, \gamma}\epsilon_{\alpha\beta\gamma}\sin{\kappa}(\bm{r})(\partial_{\alpha}\kappa) (\partial_{\beta}\eta_{1})(\partial_{\gamma}\eta_{2}), \label{eq:wind}
\end{equation}
where $\epsilon_{\alpha\beta\gamma}$ represents the antisymmetric
tensor and $\alpha$, $\beta$, and $\gamma$ are summed over $x$, $y$,
and $z$.
As long as the parameters $\kappa$, $\eta_1$, and $\eta_2$
are continuous, the winding number $W$ is an integer.
Equation~\eqref{eq:skyrmion} contains a skyrmion structure with $W = 1$.

The winding number is invariant under global spin rotation, which generates topologically equivalent states. Applying spin rotation $e^{i \sigma_y \pi / 4}$ to the state in Eq.~\eqref{eq:skyrmion}, we obtain the skyrmion state discussed in Refs.~\cite{bec1, domain, stable9, hydro}, in which component 1 has a vortex ring and its core is occupied by component 2 having $2\pi$ rotation of the phase along the ring.

\section{Numerical simulations of dynamics}
\label{s:numerical}

We consider a two-component BEC of a dilute atomic gas with mass $m$ trapped inside an external potential $V(\bm{r})$. The dynamics of the system at zero temperature under the mean-field approximation is described by the coupled Gross-Pitaevskii (GP) equations,

\begin{equation}
i\hbar\frac{\partial\Phi_j}{\partial t} = \Biggl(-\frac{\hbar^2}{2m}\nabla^2 + V(\bm{r}) + g_{jj}\vert\Phi_{j}\vert^2 + g_{jj^{'}}\vert\Phi_{j^{'}}\vert^2\Biggr)\Phi_{j}, \label{eq:gp}
\end{equation}
where $(j, j')$ = (1, 2) and (2, 1).
Here, $\Phi_{j}(\bm{r}, t)$ is normalized as $\int \vert\Phi_j\vert^2
d\bm{r}=N_j$ with $N_j$ being the number of atoms in the $j$th
component.
The coefficient $g_{jj'} = 4\pi\hbar^2a_{jj'}/m$ represents the
interaction parameter with $a_{jj'}$ being the $s$-wave scattering
length between the $j$th and $j'$th components.
In the following, we set the number of atoms in both components to be
the same, $N_{1}=N_{2}$. The inter-component interaction parameter
$g_{12}$ is assumed to be a variable that satisfies miscible condition
$g_{11}g_{22}>g^{2}_{12}$ \cite{miscible}.
For simplicity, the intra-component interactions are assumed to be
identical: $g_{11}=g_{22}\equiv g>0$.
We also assume that the system is confined in a spherically symmetric trap given by $V(\bm{r})=m\omega^{2}\bm{r}^{2}/2$, where $\omega$ is the trap frequency.

We rescale position $\bm{r}=\Tilde{\bm{r}}l_{h}$, time
$t=\omega^{-1}\tau$ and wave function $\Phi_{j}(\bm{r},
t)=\Tilde{\Phi}_{j}(\Tilde{\bm{r}}, \tau)N_{j}^{1/2}/l_{h}^{3/2}$,
where $l_h=[\hbar / (m\omega)]^{1/2}$.
Equation~\eqref{eq:gp} then takes a non-dimensional form, 
\begin{equation}
i\frac{\partial\Tilde{\Phi}_{j}}{\partial\tau} = \Biggl(-\frac{\Tilde{\nabla}^2}{2} + \Tilde{V} + \tilde{g}\vert\Tilde{\Phi}_{j}\vert^2 + \Tilde{g}_{jj'}\vert\Tilde{\Phi}_{j}\vert^2\Biggr)\Tilde{\Phi}_{j}, \label{eq:scaled}
\end{equation}
where $\Tilde{V}=\frac{\Tilde{\bm{r}}^2}{2}$, $\tilde{g} = \frac{4\pi
  N_j a}{l_h}$, and $\Tilde{g}_{jj'}=\frac{4\pi N_{j}a_{jj'}}{l_h}$
are the scaled potential and scaled interaction coefficients,
respectively.
In the following discussion, we omit the tildes from the
non-dimensional quantities.

We use the split-operator pseudo-spectral method \cite{pseudo} to
integrate the 3D GP equation numerically to obtain imaginary- and
real-time evolutions. The numerical mesh size is set to $(256)^3$ with
a spatial step size of $dx=dy=dz=0.075$, and the time step is typically
$dt=0.001$.
The numerical box is sufficiently larger than the condensate, and the periodic boundary condition imposed by the spectral method does not affect the results.

\begin{figure}[h]
\centering
\includegraphics[width=3.4in]{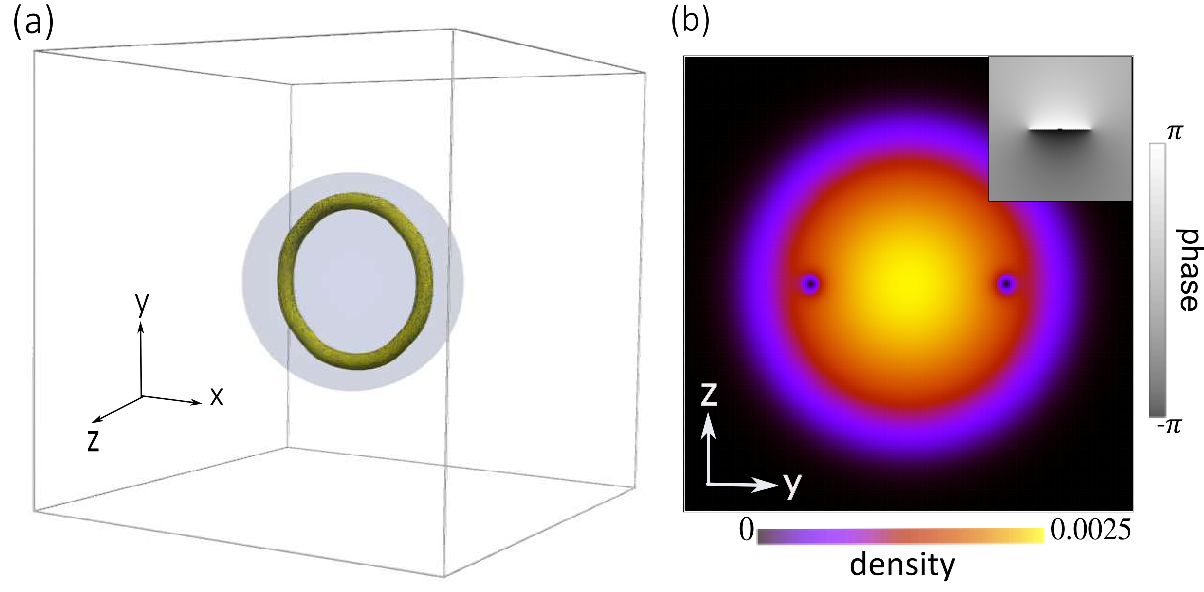}
\caption{Initial stationary state with overlapped vortex rings in
  component 1 and component 2 obtained by imaginary-time evolution of
  the GP equation with $g=5000$ and $g_{12}=0.65g$.
  The two components have the same density and phase profiles.
  (a) Isodensity surfaces at one-half the peak density, where the
  outer spherical surface is made transparent for visibility.
  (b) Cross-sectional density (main panel) and phase (inset) profiles
  of both components on the $x=0$ plane.
  We set $R_{r}=3.5l_{h}$ for preparing the stationary state.
  The size of the box in (a) is $(19.2l_h)^3$, with the origin at the
  center.
  The field of view in (b) is $19.2l_h\times19.2l_h$, and the units of
  density are $N_{j}/l^{3}_{h}$.}
\label{fig:stationary}
\end{figure}

We prepare the stationary vortex-ring state as an initial state in
which both components have the same wave function containing a
quantized vortex ring.
To generate this state numerically, we first prepare the ground state
without vortices by imaginary-time evolution.
We next imprint a circular vortex ring in both components by
multiplying the wave functions by $e^{i\Theta(\bm{r})}$, where
 \begin{equation}
\Theta(\bm{r})=\tan^{-1}\frac{z}{r_{\perp}-R_{r}} - \tan^{-1}\frac{z}{r_{\perp}+R_{r}}.
\label{eq:phase}
\end{equation}
Here, $r_{\perp}=\sqrt{x^{2}+y^{2}}$ and $R_{r}$ is the radius of the
vortex ring to be imprinted in both components.
We perform a short imaginary-time propagation (duration of 0.3) after
the phase $\Theta(\bm{r})$ is imprinted on the wave functions to
remove the excess energy of the imprinted vortices.
If the value of $R_r$ is appropriately chosen, the imaginary-time
evolution almost maintains the radii of the vortex rings and the
stationary state is achieved.
Figure~\ref{fig:stationary} shows the stationary state with overlapped
vortex rings, where $R_r=3.5l_h$ is used. In
Fig.~\ref{fig:stationary}(b), the two density holes indicate the cross
section of the vortex ring with $2\pi$ phase winding.
The two components have the same wave function and the vortex rings in
the two components are totally overlapped with each other, having the
same radius.
A small numerical perturbation is introduced to each component of the initial state before the real-time evolution to break the exact numerical symmetry and trigger the dynamical instability.

\begin{figure*}[ht!]
\centering
\includegraphics[width=\textwidth]{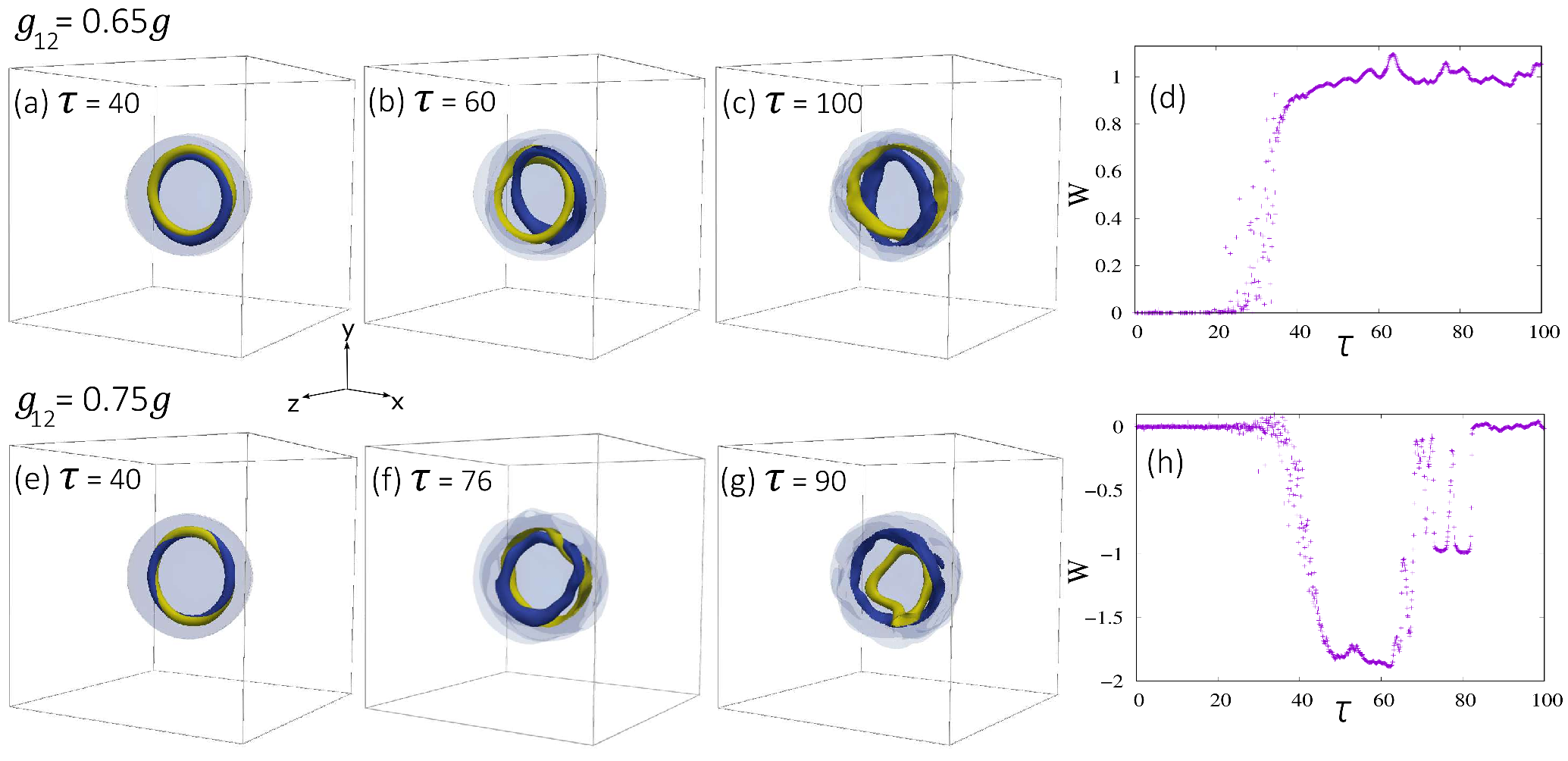}
\caption{Dynamics of spontaneous skyrmion generation obtained by the
  GP equation for $g = 5000$, where the initial state is the
  overlapped vortex rings (Fig.~\ref{fig:stationary}).
  Isodensity surfaces for component 1 (yellow or light gray) and
  component 2 (blue or dark gray) for (a)-(c) $g_{12}=0.65g$ and
  (e)-(g) $g_{12}=0.75g$.
  See the Supplemental Material for animations of the dynamics in (a)-(c) and (e)-(g). (d) and (h) Time evolution of the winding number $W$ for $g_{12}=0.65g$ and $g_{12}=0.75g$, respectively.}
\label{fig:time}
\end{figure*}

Figure~\ref{fig:time} shows the real-time evolution starting from the
stationary vortex-ring state in Fig.~\ref{fig:stationary}.
The overlapped vortex rings in the two-components are dynamically
unstable against separation.
This instability is similar to that in two overlapped vortices in two
components in 2D systems \cite{inst1, inst2}.
For a certain range of interaction parameter $g_{12}$, we find that
these coaxial vortex rings separate and become linked with each other,
which results in the skyrmion.
Figures~\ref{fig:time}(a)-\ref{fig:time}(c) show the dynamics of
spontaneous generation of a skyrmion for $g_{12}=0.65g$.
The two vortex rings already become linked immediately after
separation ($\tau \simeq 40$), and the linkage is maintained for a long
time (until $\tau \simeq 160$).
Figure~\ref{fig:time}(d) shows the time evolution of the winding
number $W$.
Since the density outside the condensate is zero and the phase is
ill-defined, the integral in Eq.~\eqref{eq:wind} is taken only for $r
< r_{\rm cutoff}=5l_{h}$.
The winding number $W$ increases from 0 to $\simeq$1 when the vortex
rings separate and become linked.
The fluctuations in the winding number during $20 \lesssim \tau
\lesssim 40$ are due to the ill-defined phases near the separating
vortex cores, and those for $\tau \gtrsim 40$ are due to the cutoff
$r_{\rm cutoff}$.
At $\tau \simeq 160$, the link between the vortex rings breaks and the
winding number returns to $W \simeq 0$ (data not shown).
The skyrmion with $W = 1$ or $-1$ is obtained randomly depending on
the initial random noise.
In the case of $g_{12}=0.75g$, spontaneous generation of a skyrmion
with double-winding linkages is obtained at $\tau\simeq 40$, as shown
in Fig.~\ref{fig:time}(e), which has the winding number $W \simeq -2$.
After $\tau\simeq 65$, it reduces to the single-winding linkage
[Fig.~\ref{fig:time}(f)], followed by the leapfrog dynamics at
$\tau\simeq 90$ [Fig.~\ref{fig:time}(g)].
For $g_{12}=0.60g$, no skyrmion is generated and only the leapfrog
dynamics is observed (Fig.~\ref{fig:0.60}).
The results show that the separated vortex rings are not linked with
each other and that the axial symmetry is retained for a long time.
The winding number \textit{W} for this case remains zero in the time evolution.

We performed systematic simulations using the GP equation for various
values of $g_{12}$ and found that the skyrmion with $|W| \simeq 1$ is
generated for $0.62g \lesssim\ g_{12}\lesssim 0.70g$.
The leapfrog dynamics of the vortex rings are obtained for $g_{12}
\lesssim 0.61g$.
For $0.74g \lesssim g_{12} \lesssim 0.77g$, we obtain the skyrmion
with $|W| \simeq 2$ with a short lifetime, followed by the leapfrog dynamics.

\section{Bogoliubov Analysis}
\label{s:Bogo}

We here use the Bogoliubov analysis to investigate the dynamical stability of the system.
We separate the wave function as 
\begin{equation}
\Phi_{j}(\bm{r}, \tau) = [\psi_{j}(r_{\perp}, z) + \delta\Phi_{j}(\bm{r}, \tau)]e^{-i\mu_{j}\tau}, \label{eq:sep}
\end{equation}
where $\psi_{j}$ is the stationary state, $\delta\Phi_{j}$ is the small deviation, and $\mu_{j}$ is the chemical potential. Substituting Eq.~\eqref{eq:sep} into the GP equation~\eqref{eq:scaled} and neglecting the second- and third-order terms in $\delta\Phi_{j}$, we obtain $(j \neq j')$
\begin{equation}
  \begin{aligned}
&i\frac{\partial{(\delta\Phi_{j})}}{\partial{\tau}} = \Biggl(-\frac{\nabla^2}{2} + V - \mu_{j}\Biggr)\delta\Phi_{j} + g_{jj}(2\vert\psi_{j}\vert^2\delta\Phi_{j} 
\\
&+ \psi^{2}_{j}\delta\Phi^{*}_{j}) + g_{jj'}(\vert\psi_{j'}\vert^2\delta\Phi_{j} + \psi^{*}_{j'}\psi_{j}\delta\Phi_{j'} + \psi_{j'}\psi_{j}\delta\Phi^{*}_{j'}).
  \end{aligned}\label{eq:subs}
\end{equation}
We write the deviation $\delta\Phi_j$ as
\begin{equation}
\delta\Phi_{j}(\bm{r}, \tau) = u_{j}(r_{\perp}, z)e^{i\ell\phi}e^{-i\lambda\tau} + v_{j}^{*}(r_{\perp}, z)e^{-i\ell\phi}e^{i\lambda^{*}\tau},
\label{eq:fluctuation}
\vspace{12pt}
\end{equation}
where an integer $\ell$ denotes the angular momentum of the excitation. Substituting Eq.~\eqref{eq:fluctuation} into Eq.~\eqref{eq:subs}, we obtain the Bogoliubov-de Gennes equation $(j \neq j')$,
\begin{subequations}
\begin{equation}
  \begin{aligned}
&\bigg[-\frac{1}{2}\biggl({\nabla}^{2}_{\perp} + {\nabla}^{2}_{z} - \frac{\ell^2}{r^{2}_{\perp}}\biggr) + V - \mu_{j}\bigg]u_{j} + g_{jj}(2\vert\psi_{j}\vert^{2}u_{j} 
\\
&+ \psi^{2}_{j}v_{j}) + g_{jj'}(\vert\psi_{j'}\vert^{2}u_{j} \delta\psi_{j} + \psi^{*}_{j'}\psi_{j}u_{j'} + \psi_{j'}\psi_{j}v_{j'}) = \lambda u_{j},
  \end{aligned}
\end{equation}
\begin{equation}
  \begin{aligned}
&\bigg[-\frac{1}{2}\biggl({\nabla}^{2}_{\perp} + {\nabla}^{2}_{z} - \frac{\ell^2}{r^{2}_{\perp}}\biggr) + V - \mu_{j}\bigg]v_{j} + g_{jj}(2\vert\psi_{j}\vert^{2}v_{j} 
\\
&+ \psi^{*2}_{j}u_{j}) + g_{jj'}(\vert\psi_{j'}\vert^{2}v_{j} + \psi_{j'}\psi^{*}_{j}v_{j'} + \psi^{*}_{j'}\psi^{*}_{j}u_{j'}) = -\lambda v_{j}. 
  \end{aligned}
\end{equation}\label{eq:bdgs}
\end{subequations}

Numerically, we obtain the stationary state $\psi_j(r_\perp, z)$
containing vortices using the imaginary-time evolution of the GP
equation, in a manner similar to Fig.~\ref{fig:stationary}, where
$r_\perp$ and $z$ are discretized with the step size $dr_\perp = dz =
0.075$.
We next refine the stationary state using the Newton-Raphson method
[Fig.~\ref{fig:bogo}(a)].
Using this stationary state, we calculate the eigenvalues and
eigenfunctions of Eq.~\eqref{eq:bdgs} using the Lanczos method.
The existence of at least one eigensolution with a positive imaginary
part (Im$\lambda>0$) indicates dynamic instability because the
corresponding excitation mode exhibits exponential growth in time.

An example of the eigenfunctions $u_j$ and $v_j$ for a dynamically
unstable mode is shown in Figs.~\ref{fig:bogo}(b)-\ref{fig:bogo}(e).
Since Eq.~(\ref{eq:bdgs}) does not depend on the sign of $\ell$, we
show only the results for $\ell\geq0$.
The eigenfunctions $u_j$ are localized near the vortex core
[Figs.~\ref{fig:bogo}(b) and \ref{fig:bogo}(c)], whereas $v_j$ are
distributed [Figs.~\ref{fig:bogo}(d) and \ref{fig:bogo}(e)].
They satisfy $\int d\bm{r} \sum_{j=1}^2 (|u_j|^2 - |v_j|^2) = 0$. We
note that the localized functions satisfy $u_1 = -u_2$, which leads to
vortex separation, as discussed below.
The eigenfunctions for other unstable modes have similar shapes.
There is also a dynamically unstable mode in which $u_j$ and $v_j$ are
exchanged, since $(\lambda, u_{j}, v_{j})$ and $(-\lambda^*, v^*_{j},
u^*_{j})$ are conjugate solutions.

\begin{figure}[t]
\centering
\includegraphics[width=3.4in]{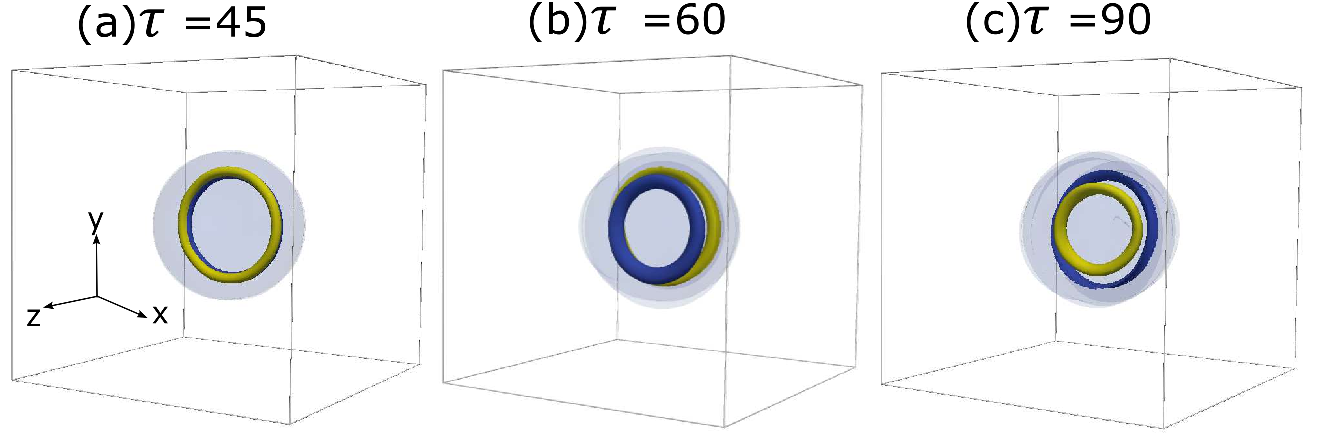}
\caption{Leapfrog dynamics obtained by solving the GP equation for $g = 5000$ and $g_{12}=0.60g$. (a) $\tau=45$, (b) $\tau=60$, and (c) $\tau=90$. Isodensity surfaces for both components are shown. See the Supplemental Material for an animation of the dynamics.}
\label{fig:0.60}
\end{figure}

The integer $\ell$ for the unstable mode corresponds to the winding
number $W$ for the generated state, which can be understood as
follows.
Suppose that the core of the vortex ring is located at $r_\perp = r_0$
and $z = 0$, as shown in Fig.~\ref{fig:bogo}(a).
We focus on the region very close to the core of the vortex ring,
where $u_j$ are dominant in the excitation.
Approximating that $u_j$ are uniform near the vortex core, we can
express the time evolution of the wave function as
 \begin{equation}
\psi_{j}  \propto r_{\perp}-r_{0} + iz + C_{j}(t)e^{i\ell\phi},
\label{eq:last}
\end{equation}
where $C_{j}$ are proportional to the values of $u_j$ at the vortex core. Without loss of generality, $C_j$ can be taken to be real, and Eq.~\eqref{eq:last} becomes
 \begin{equation}
\psi_{j}  \propto r_{\perp}-(r_0 - C_j \cos\ell\phi) + i (z + C_j \sin\ell\phi).
\label{eq:lastpro}
\end{equation}
Thus, the position of the vortex core is displaced by $(-C_j
\cos\ell\phi, -C_j \sin\ell\phi)$; i.e., the displaced vortex ring
winds $\ell$ times around the original ring from $\phi = 0$ to $2\pi$,
which corresponds to the winding number $W = \ell$ for the skyrmion.
Since $u_1 = -u_2$ (i.e., $C_1 = -C_2$), the vortices in the two
components separate from each other.
\begin{figure}[h]
\centering
\includegraphics[width=3.4in]{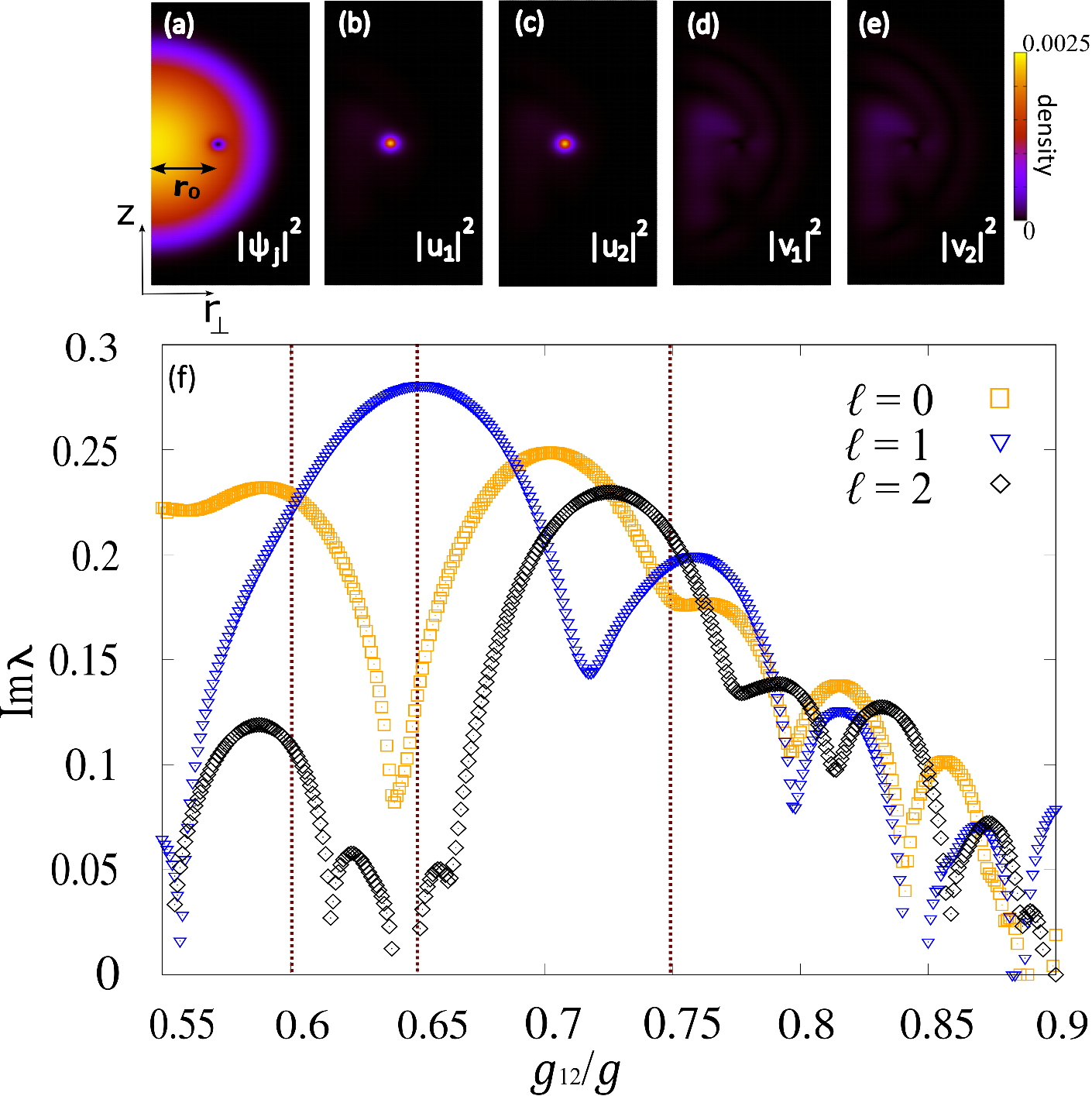}
\caption{Bogoliubov analysis for $g = 5000$.
  (a) Density profile for the stationary state $\psi_{j}$.
  (b), (c), (d), and (e) Density profiles for dynamically unstable
  modes $u_1$, $u_2$, $v_1$, and $v_2$, respectively, for $g_{12} =
  0.65g$ and $\ell=1$.
  (f) Imaginary part of the Bogoliubov excitation frequency
  Im$\lambda$ for eigenmodes with $\ell$ = 0, 1, and 2 plotted as a
  function of $g_{12}/g$.
  The vertical dotted lines indicate $g_{12}/g=0.60$, $g_{12}/g=0.65$, and $g_{12}/g=0.75$, corresponding to the parameters used in Fig.~\ref{fig:time} and \ref{fig:0.60}.}
\label{fig:bogo}
\end{figure}

Figure~\ref{fig:bogo}(f) shows the imaginary part of the excitation
frequency, Im$\lambda$, for $\ell= 0$, 1, and 2 as a function of
$g_{12}/g$.
The system is dynamically unstable in the entire range shown, $0.55
\lesssim g_{12}/g \lesssim 0.9$, since there are excitation
frequencies with a positive imaginary part for any values of
$g_{12}/g$.
For $g_{12} / g \simeq 0.6$, 0.65, and 0.75, the excitation frequency
$\lambda$ has the largest imaginary part for $\ell = 0$, 1, and 2 (see
the plots at the vertical dotted lines in Fig.~\ref{fig:bogo}(f)).
This result is consistent with the dynamics in Figs.~\ref{fig:time}
and \ref{fig:0.60}, where the generated states have winding numbers
$|W| = 0$, 1, and 2 for $g_{12} / g \simeq 0.6$, 0.65, and 0.75,
respectively.
Although Im$\lambda$ also becomes nonzero for the modes of $\ell \geq
3$, the values of Im$\lambda$ for $\ell \geq 3$ are always smaller
than those for $\ell \leq 2$.
We have confirmed that the results are almost unchanged when the
spatial step size is doubled, which indicates that the numerical error
due to the spatial resolution is unimportant.

\section{Experimental Proposal}
\label{s:exp}

We consider a two-component BEC of $^{87}{\rm Rb}$ atoms confined in
an isotropic harmonic potential $V = m\omega^{2}(x^2 + y^2 + z^2)/2$
with a trap frequency $\omega$ = 2$\pi\times$100 Hz.
In this case, the units of length and time are $l_h = 1.08 \mu{\rm m}$ and
$\omega^{-1} = 1.59$ ms, respectively.
In Secs. III and IV, for simplicity, we assumed $g_{11} = g_{22}$,
which can be realized by, e.g., using the hyperfine states $\ket{F,
  m_F}$ = $\ket{1, 1}$ and $\ket{1, -1}$ for the two components.
The s-wave scattering lengths for these states are
$a_{11}=a_{22}=100.4a_0$ \cite{bohr1, bohr2}, where $a_0$ is the Bohr
radius.
The interaction parameter $g = 5000$ corresponds to $N_{1}=N_{2}\simeq8\times10^{4}$.

The initial state in which the two components contain the overlapped
vortex rings (Fig.~\ref{fig:stationary}) can be prepared as follows.
First, a single-component BEC is created and a stationary vortex ring
is produced in the BEC \cite{bec1, imprint1, gaussian}.
Next, one-half of one component is transferred to the other component
through the Rabi oscillation using a $\pi/2$ pulse, which results in
the overlapped vortex rings in the two-component BEC.

We have numerically confirmed through the GP simulations that similar
results can be obtained for the hyperfine states $\ket{1, 1}$ and
$\ket{1, 0}$ of $^{87}{\rm Rb}$ atoms, where the intracomponent
scattering lengths $a_{11}=100.4a_0$ and $a_{22}=100.86a_0$ are
different.
In the numerical simulation, we prepared a stationary vortex-ring
state for a BEC of the $\ket{1, 1}$ state and then one-half of the
atoms were transferred to the $\ket{1, 0}$ state.
We then carried out the real-time evolution in a manner similar to
Fig.~\ref{fig:time}.
We found that the skyrmion is generated for $0.77 g_{11} \lesssim
g_{12} \lesssim 0.79 g_{11}$.

\section{Conclusions}
\label{s:conc}

We have shown that a 3D skyrmion is spontaneously generated in a
two-component miscible BEC trapped in a harmonic potential, in which
intercomponent interaction $g_{12}$ is controlled.
For a certain range of the intercomponent interaction parameter, the
initially overlapped vortex rings exhibit dynamical instability and
separate into two linked vortex rings (Fig.~\ref{fig:time}).
We carried out the Bogoliubov analysis and numerically obtained the
dynamically unstable spectrum (Fig.~\ref{fig:bogo}).
This phenomenon can be realized in a realistic experimental system
using, e.g., a two-component BEC with a feasible number of atoms
($\sim10^{5}$ atoms of $^{87}{\rm Rb}$).

\section*{ACKNOWLEDGMENT}

We thank K. Miyamoto for participating in the early stage of this work. This work was supported by JSPS KAKENHI Grant No. JP23K03276.

\bibliographystyle{plain}

\thebibliography{references}
\bibitem{skyrme} T. H. R Skyrme, A non-linear field theory, Proc. R. Soc. London Ser. A \textbf{260}, 127 (1961); A unified field theory of mesons and baryons, Nucl. Phys. \textbf{31}, 556 (1962).

\bibitem{m1}  A. N. Bogdanov and U. K. Rößler, Chiral Symmetry Breaking in Magnetic Thin Films and Multilayers, Phys. Rev. Lett. \textbf{87}, 037203 (2001); U. K. Rößler, A. N. Bogdanov, and C. Pfleiderer, Spontaneous skyrmion ground states in magnetic metals, Nature (London) \textbf{442}, 797 (2006).

\bibitem{crys} A. N. Bogdanov, U. K. Rößler, and A. A. Shestakov, Skyrmions in nematic liquid crystals, Phys. Rev. E \textbf{67}, 016602 (2003).

\bibitem{hall1} D.-H. Lee and C. L. Kane, Boson-vortex-skyrmion duality, spin-singlet fractional quantum Hall effect, and spin-1/2 anyon superconductivity, Phys. Rev. Lett. \textbf{64}, 1313 (1990).

\bibitem{hall2} S. L. Sondhi, A. Karlhede, S. A. Kivelson, and E. H. Rezayi, Skyrmions and the crossover from the integer to fractional quantum Hall effect at small Zeeman energies, Phys. Rev. B \textbf{47}, 16419 (1993).

\bibitem{hall3} A. Schmeller, J. P. Eisenstein, L. N. Pfeiffer, and K. W. West, Evidence for Skyrmions and single spin flips in the integer quantized Hall effect, Phys. Rev. Lett. \textbf{75}, 4290 (1995).

\bibitem{super1} J. Garaud, J. Calstr\"om, and E. Babaev, Topological solitons in three-band superconductors with broken time reversal symmetry, Phys. Rev. Lett. \textbf{107}, 197001 (2011).

\bibitem{super2} S. S. Pershoguba, S. Nakosai, and A. V. Balatsky, Skyrmion-induced bound states in a superconductor, Phys. Rev. B \textbf{94}, 064513 (2016).

\bibitem{solid} J. Koralek, C. P. Weber, J. Orenstein, B. A. Bernevig, Shou-Cheng Zhang, S. Mack, D. D. Awschalom, Emergence of the persistent spin helix in semiconductor quantum wells, Nature (London) \textbf{458}, 610 (2009).

\bibitem{acous}  H. Ge, X.-Y. Xu, L. Liu, R. Xu, Z.-K. Lin, S.-Y. Yu, M. Bao,
J.-H. Jiang, M.-H. Lu, and Y.-F. Chen, Observation of Acoustic Skyrmions, Phys. Rev. Lett. \textbf{127}, 144502 (2021).

\bibitem{flu1} N. D Mermin and T.-L. Ho, Circulation and Angular Momentum in the \textit{A} Phase of Superfluid Helium-3 Phys. Rev. Lett. \textbf{36}, 594 (1976).

\bibitem{flu2} P. W. Anderson and G. Toulouse, Phase Slippage without Vortex Cores: Vortex Textures in Superfluid $^{3}{\rm He}$, Phys. Rev. Lett. \textbf{38}, 508 (1977).

\bibitem{bec1} J. Ruostekoski and J. R. Anglin, Creating Vortex Rings and Three-Dimensional Skyrmions in Bose-Einstein Condensates, Phys. Rev. Lett. \textbf{86}, 3934 (2001).

\bibitem{bec2} U. Al Khawaja and H. Stoof, Skyrmions in a ferromagnetic Bose-Einstein condensate, Nature (London) \textbf{411}, 918 (2001).

\bibitem{soliton1} S. Burger, K. Bongs, S. Dettmer, W. Ertmer, K. Sengstock, A. Sanpera, G. V. Shlyapnikov, and M. Lewenstein, Dark solitons in Bose-Einstein condensates, Phys. Rev. Lett. \textbf{83}, 5198 (1999).

\bibitem{vortices1} M. R. Matthews, B. P. Anderson, P. C. Haljan, D. S. Hall, C. E. Wieman, and E. A. Cornell, Vortices in a Bose-Einstein condensate, Phys. Rev. Lett. \textbf{83}, 2498 (1999).

\bibitem{vortices2} A. E. Leanhardt, Y. Shin, D. Kielpinski, D. E. Pritchard, and W. Ketterle, Coreless vortex formation in a spinor Bose-Einstein condensate, Phys. Rev. Lett. \textbf{90}, 140403 (2003).

\bibitem{mono1} H. T. C. Stoof, E. Vliegen, and U. Al Khawaja, Monopoles in an antiferromagnetic Bose-Einstein condensate, Phys. Rev. Lett. \textbf{87}, 120407 (2001).

\bibitem{mono2} V. Pietilä and M. Möttönen, Non-abelian magnetic monopole in a Bose-Einstein condensate, Phys. Rev. Lett. 102, 080403 (2009); Creation of dirac monopoles in spinor Bose-Einstein condensates, Phys. Rev. Lett. \textbf{103}, 030401 (2009).

\bibitem{mono3} M. W. Ray, E. Ruokokoski, K. Tiurev, M. Möttönen, and D. S. Hall, Observation of isolated monopoles in a quantum field, Science \textbf{348}, 544 (2015).

\bibitem{knots1} Y. Kawaguchi, M. Nitta, and M. Ueda, Knots in a spinor Bose-Einstein condensate, Phys. Rev. Lett. \textbf{100}, 180403 (2008).

\bibitem{knots2} D. S. Hall, M. W. Ray, K. Tiurev, E. Ruokokoski, A. H. Gheorghe and M. Möttönen, Tying quantum knots, Nature Phys. \textbf{12}, 478 (2016).

\bibitem{knots3} T. Annala, T. Mikkonen, M. Möttönen, Optical
  induction of monopoles, knots, and skyrmions in quantum gases,
Phys. Rev. Research \textbf{6}, 043013 (2024).

\bibitem{raman} L. S. Leslie, A. Hansen, K. C. Wright, B. M. Deutsch, and N. P. Bigelow, Creation and detection of skyrmions in a Bose-Einstein Condensate, Phys. Rev. Lett. \textbf{103}, 250401 (2009).

\bibitem{rot} J.-Y. Choi, W. J. Kwon, and Y.-I. Shin, Observation of topologically stable 2D skyrmions in an antiferromagnetic spinor Bose-Einstein Condensate, Phys. Rev. Lett. \textbf{108}, 035301 (2012).

\bibitem{spin-1} W. Lee, A. H. Gheorghe, K. Tiurev, T. Ollikainen, M. Möttönen, and D. S. Hall, Synthetic electromagnetic knot in a three-dimensional skyrmion, Sci. Adv. \textbf{4}, 3820 (2018).
 
\bibitem{spin-2} K. Tiurev, T. Ollikainen, P. Kuopanportti, M. Nakahara, D. S. Hall, and M. Möttönen, Three-dimensional skyrmions in spin-2 Bose–Einstein condensates, New J. Phys. \textbf{20}, 055011 (2018).

\bibitem{zamora} R. Zamora-Zamora and V. Romero-Rochin, Skyrmions with arbitrary topological charges in spinor Bose–Einstein condensates, J. Phys. B \textbf{51}, 045301 (2018).

\bibitem{q} H.-B. Luo, L. Li, and W.-M. Liu, Three-dimensional skyrmions with arbitrary topological number in a ferromagnetic spin-1 Bose-Einstein condensate, Sci. Rep. \textbf{9}, 18804 (2019).

\bibitem{imprint} Z. Chen, S. X. Hu, and N. P. Bigelow, Imprinting a three-dimensional skyrmion in a Bose–Einstein condensate via a Raman process, J. Low Temp. Phys. \textbf{208}, 172 (2022).

\bibitem{coupling1} T. Kawakami, T. Mizushima, M. Nitta, and K. Machida, Stable skyrmions in $SU(2)$ gauged Bose-Einstein condensates, Phys. Rev. Lett. \textbf{109}, 015301 (2012).

\bibitem{coupling2} G. Chen, T. Li, and Y. Zhang, Symmetry-protected skyrmions in three-dimensional spin-orbit-coupled Bose gases, Phys. Rev. A \textbf{91}, 053624 (2015).

\bibitem{domain} M. Nitta, K. Kasamatsu, M. Tsubota, and H. Takeuchi, Creating vortons and three-dimensional skyrmions from domain-wall annihilation with stretched vortices in Bose-Einstein condensates, Phys. Rev. A \textbf{85}, 053639 (2012).

\bibitem{capillary} K. Sasaki, N. Suzuki, and H. Saito, Capillary instability in a two-component Bose-Einstein condensate, Phys. Rev. A \textbf{83}, 053606 (2011).

\bibitem{optical} C. D. Parmee, M. R. Dennis and J. Ruostekoski, Optical excitations of Skyrmions, knotted solitons, and defects in atoms, Commun. Phys. \textbf{5}, 54 (2022).

\bibitem{hydro} K. Sakaguchi, K. Jimbo, and H. Saito, Hydrodynamic generation of skyrmions in a two-component Bose-Einstein condensate, Phys. Rev. A \textbf{105}, 013312 (2022).

\bibitem{stoof} U. Al Khawaja and H. T. C. Stoof, Skyrmion physics in Bose-Einstein ferromagnets, Phys. Rev. A \textbf{64}, 043612 (2001).

\bibitem{stable1} R. A. Battye, N. R. Cooper, and P. M. Sutcliffe, Stable skyrmions in two-component Bose-Einstein condensates, Phys. Rev. Lett. \textbf{88}, 080401 (2002).

\bibitem{lifetime} Y. Zhang, W.-D. Li, L. Li, and H. J. W. Müller-Kirsten, Exact calculation of the skyrmion lifetime in a ferromagnetic Bose-Einstein condensate, Phys. Rev. A \textbf{66}, 043622 (2002).

\bibitem{Simula} T. P. Simula, J. A. M. Huhtamäki, M. Takahashi, T. Mizushima, and K. Machida, Rotating Dipolar Spin-1 Bose–Einstein Condensates, J. Phys. Soc. Jpn. \textbf{80}, 013001 (2011)

\bibitem{stable2} H. Zhai, W. Q. Chen, Z. Xu, and L. Chang, Skyrmion excitation in a two-dimensional spinor Bose-Einstein condensate, Phys. Rev. A \textbf{68}, 043602 (2003).

\bibitem{stable3} C. M. Savage and J. Ruostekoski, Energetically stable particlelike skyrmions in a trapped Bose-Einstein condensate, Phys. Rev. Lett. \textbf{91}, 010403 (2003).

\bibitem{stable4} J. Ruostekoski, Stable particlelike solitons with multiply quantized vortex lines in Bose-Einstein condensates, Phys. Rev. A \textbf{70}, 041601(R) (2004).

\bibitem{stable5} S. Wüster, T. E. Argue, and C. M. Savage, Numerical study of the stability of skyrmions in Bose-Einstein condensates, Phys. Rev. A \textbf{72}, 043616 (2005).

\bibitem{stable6} I. F. Herbut and M. Oshikawa, Stable Skyrmions in spinor condensates, Phys. Rev. Lett. \textbf{97}, 080403 (2006).

\bibitem{stable7} A. Tokuno, Y. Mitamura, M. Oshikawa, and I. F. Herbut, Skyrmion in spinor condensates and its stability in trap potentials, Phys. Rev. A \textbf{79}, 053626 (2009).

\bibitem{stable8} H. M. Price and N. R. Cooper, Skyrmion-antiskyrmion pairs in ultracold atomic gases, Phys. Rev. A \textbf{83}, 061605(R) (2011).

\bibitem{stable9} T. Kaneda and H. Saito, Collision dynamics of skyrmions in a two-component Bose-Einstein condensate, Phys. Rev. A \textbf{93}, 033611 (2016).

\bibitem{stable10} A. Samoilenka, F. N. Rybakov, and E. Babaev, Synthetic nuclear Skyrme matter in imbalanced Fermi superfluids with a multicomponent order parameter, Phys. Rev. A \textbf{101}, 013614 (2020).

\bibitem{top} N. Manton and P. Sutcliffe, \textit{Topological Solitons}, (Cambridge University Press, Cambridge, 2004).

\bibitem{bar1} S. B. Gudnason and M. Nitta, Linking number of vortices as baryon number, Phys. Rev. D \textbf{101}, 065011 (2020). 

\bibitem{miscible} C. J. Pethick and H. Smith, \textit{Bose-Einstein Condensation in Dilute Gases}, 2nd ed. (Cambridge University Press, Cambridge, U.K., 2007).

\bibitem{pseudo} W. H. Press, S. A. Teukolsky, W. T. Vetterling, and B. P. Flannery, \textit{Numerical Recipes}, 3rd ed. (Cambridge University Press, Cambridge, 2007).

\bibitem{inst1} P. Kuopanportti, S. Bandyopadhyay, A. Roy, and D. Angom, Splitting of singly and doubly quantized composite vortices in two-component Bose-Einstein condensates, Phys. Rev. A \textbf{100}, 033615 (2019).

\bibitem{inst2} J. Han, K. Kasamatsu, and M. Tsubota, Dynamics of two quantized vortices belonging to different components of binary Bose-Einstein condensates in a circular box potential, J. Phys. Soc. Jpn. \textbf{91}, 024401 (2022).

\bibitem{bohr1} E. G. M. van Kempen, S. J. J. M. F. Kokkelmans, D. J. Heinzen, and B. J. Verhaar, Interisotope determination of ultracold rubidium interactions from three high-precision experiments, Phys. Rev. Lett. \textbf{88}, 093201 (2002).

\bibitem{bohr2} A. Widera, F. Gerbier, S. Fölling, T. Gericke, O. Mandel, and I. Bloch, Precision measurement of spin-dependent interaction strengths for spin-1 and spin-2 $^{87}{\rm Rb}$ atoms New J. Phys. \textbf{8}, 152 (2006).

\bibitem{imprint1} J. Ruostekoski and Z. Dutton, Engineering vortex rings and systems for controlled studies of vortex interactions in Bose-Einstein condensates, Phys. Rev. A \textbf{72}, 063626 (2005).

\bibitem{gaussian} H. Saito, Creation and Manipulation of Quantized Vortices in Bose–Einstein Condensates Using Reinforcement Learning, J. Phys. Soc. Jpn. \textbf{89}, 074006 (2020).

\end{document}